\documentclass[twocolumn,showpacs,preprintnumbers,amsmath,amssymb,superscriptaddress]{revtex4}
\usepackage{epsfig,amssymb,amsmath}
\usepackage{endnotes,rotating}
\usepackage[maccyr]{inputenc}
\usepackage[T2A]{fontenc}
\usepackage[english,russian]{babel}
\usepackage{bm}%
\usepackage{xcolor}%
\allowdisplaybreaks[4]

\begin{document}
\author{D. Kovalenko}
\email{deniskov@students.cs.ubc.ca}
 \affiliation{University of British Columbia, Academic
2366 Main Mall, Vancouver, BC, Canada}
\title{One class of integrals evaluation in magnet soliton theory II.}
\author{Zhmudsky A.A.}
\email{ozhmudsky@physics.ucf.edu}
 \affiliation{Department of Physics, University of Central Florida, 4000 Central Florida Blvd. Orlando, FL, 32816}
\begin{abstract} 
The paper proposes an approximate expression for calculating
very complex one-dimensional integrals depending on the parameter $a$.
These integrals often occur in computational problems
theory of magnetic solitons.

The resulting analytical expressions are verified by numerical calculation.
The comparison shows that the relative accuracy of the approximating expression tends to 
zero for large and small values of the parameter $a$. In a small region near $a = 0.119$, the error does 
not exceed five percent, and for parameter values near $18.0$
the relative error does not exceed seven percent. It turned out that the proposed method of constructing
approximating expressions can be generalized to much more complex integrals.

A program that compares an approximate expression with an exact numerical value
 written in C and tested on examples that allow analytical solutions.
\end{abstract} 

\maketitle
\section{Introduction}
Nonlinear excitations (topological solitons \cite{KIK}) play an important
role in the physics of low-dimensional magnets \cite{BI,IvK}. They
contribute greatly to a heat capacity, susceptibility, scattering
cross-section and other physical characteristics of magnets. In particular,
for two-dimensional (2D) magnets with discrete degeneracy it is important
to take into account the localized stable (with quite-long life time) 2D
solitons \cite{BI,IvK}. According to the experiments \cite{W1983,W1986}, these
solitons determine the relaxation of magnetic disturbance and can produce
peaks in the response functions.

The traditional model describes the state of the magnet in units
magnetization vector $\vec m$, $\vec m^2=1$ with energy function in form
\begin{equation}
W=\int\limits_{}^{}d^2x\left\{A(\nabla\vec m)^2+w_0(\vec m)\right\},
\label{resid:1}
\end{equation}
where A is the inhomogeneous exchange constant, and $w_0(\vec m)$ is
anisotropy energy.

In the anisotropic case, the solution is multidimensional, so
the soliton structure is determined by a system of equations in partial 
derivatives. There are no general methods for finding localized
solutions of such equations and stability analysis of the solution.
For this reason, the direct variational method is often used. Consequently,
the choice of trial function plays a key role in such an analysis
\cite{ZyI,IZ,Steph}. In most cases, the choice of trial
function in the form:
\begin{gather} 
tg {\theta \over 2}={R \over r}\exp \left(-\frac rb \right)
(1+C_1\cos2\chi), \\
\varphi=\chi+C_2\sin2\chi+\varphi_0,
\label{resid:2}\end{gather}
is quite successful. In this formula:
  $r$, $\chi$ are polar coordinates in the magnetic plane, $R$, $a$,
$C_1$, $C_2$ and $\varphi_0$ are variable parameters.

In the papers \cite{ZyI,IZ,Steph} Newton's iterative method for solving a system of non-linear algebraic equations \cite{teuk} 
was used to find variable parameters that provide a minimum of energy. The numerical algorithm indicated above leads to the 
necessity of multiple calculation of two-dimensional integrals (over the angle $\chi$ and the radius $r$). Experience shows 
that the calculation process can be significantly accelerated (and the accuracy of calculations improved) if an analytic-numerical 
method for calculating integrals over the radius is prepared in advance. The typical shape of these
integrals:
\begin{equation} \int\limits_0^{\infty}{f(r)dr\over\left[(r\exp(r))^2+b^2\right]^n}
\label{resid:3}\end{equation}
where $n$ is an integer, $f(r) = \exp(r), r\exp(r), \\ (1 + r)\exp(r), r^2\exp(r)$ and etc.
   Note that $a$ is an essential parameter - there is no replacement that could eliminate it. Below we consider the case $n=1$. 
   On the one hand, it is almost always possible to reduce the calculation for $n>1$ to the sum of integrals for $n=1$. On the other side
it is not difficult to modify the calculation scheme for the case $n>1$.   

In a previous work \cite{One} it was shown that
with an appropriate choice of the integration contour in the complex plane, followed by numerical calculation of the 
sums of residues of the integrand, one can find values of the integral of the type (\ref{resid:3}) with a given accuracy. 
However, in this case it is impossible to write an analytic expression for the roots in the complex plane. This leads 
to the need to take into account the number of poles of the integrand that is not specified in advance.
 
In this paper, we propose a simple analytical expression that does not require the calculation of a predetermined 
number of poles and allows to find the value of the integral with an accuracy of no more than a few percents.

As an example, consider two integrals:
\begin{gather} 
 I(a) = \int\limits_0^{\infty} \frac {x\exp(x)dx}{(x\exp(x))^2+a^2},   \label{sma} \\
 J(a) = \int\limits_0^{\infty} \frac {\exp(x) dx}{(x\exp(x))^2+a^2}   \label{smb}
\end{gather} 
It is easy to show that their sum is
\begin{gather}
 I(a) +  J(a) = \frac \pi{2a}  \quad   
\end{gather}
This equality will be useful for controlling the accuracy of numerical integration.  
\section{For small values  of the parameter $a$}\label{small} 
For parameter values $a\to 0$, one can obtain a good approximation for integrals (\ref{sma},\ref{smb}).  
Let us introduce a new variable $u = x\exp(x)$. The solution of this equation with respect to $x$ is expressed 
in terms of the Lambert function $x=W_0(u)$ \cite{Lam, Eul, Dub}. Integrals (\ref{sma},\ref{smb}) become:  
\begin{gather}
 I(a) = \int\limits_0^{\infty} \frac {W_0(u)}{1+W_0(u)} \frac { du}{u^2+a^2},  
 J(a) = \int\limits_0^{\infty} \frac 1{1+W_0(u)} \frac { du}{u^2+a^2}   
\end{gather} 
where $W_0(u)$ is one of the principal branches of the Lambert function.  

Let us rewrite the integrals in a form:
\begin{gather}
 I(a) =  \frac \pi {2 a} \int\limits_0^{\infty} \frac {W_0(u)}{1+W_0(u)} \frac 2\pi \frac {a du}{u^2+a^2},    \\
 J(a) =  \frac \pi {2 a} \int\limits_0^{\infty} \frac 1{1+W_0(u)} \frac 2\pi \frac {a du}{u^2+a^2}   
\end{gather} 
 The following multiplier:   
\begin{gather}
\frac 2\pi \frac {a}{u^2+a^2}  \qquad 
\end{gather} 
when $a\to 0$  can be treated as delta function normalized from $0$ to $\infty$.  

So, in the limit  $a\to 0$ we  get:  
\begin{gather}
 I(a) =  \frac \pi {2 a}  \left[\frac {W_0(a)}{1+W_0(a)}\right]_{a=0} = 0 ,    \label{i0} \\
 J(a) =  \frac \pi {2 a}   \left[\frac 1{1+W_0(a)}\right]_{a=0} = \frac \pi 2   \label{j0}
\end{gather}  
In formulas (\ref{i0},\ref{j0}) we take into account that $\lim\limits_{u\to 0} W_0(u) = u$.

Further, we show that these expressions describe well the integrals (\ref{sma}, \ref{smb}) not only for small values of $a$, but also for any $a > 0$.
Also, comparison with the results of numerical calculations shows that the relative error of these formulas does not exceed 
$7 \%$ for any values of $a > 0$.
\section{Approximate analytic expressions}\label{approx}
To find approximate analytic expressions, we integrate the first of the integrals (\ref{sma}) with respect to the parameter {\bf a}:
\begin{gather*}
\int I(a) da = \int \int\limits_0^{\infty} \frac {x\exp(x)dx}{(x\exp(x))^2+a^2} da = \\
 \int\limits_0^{\infty} \arctg \displaystyle\left(\frac a{x\exp(x)}\right) dx
\end{gather*} 
The following figure shows the graph of the function $\arctg\displaystyle\left( \frac a{x\exp(x)} \right)$ for parameter $a = 10, 10^2, 10^3, 10^4, 10 ^5$
\begin{figure}[h]
\includegraphics[width=76mm]{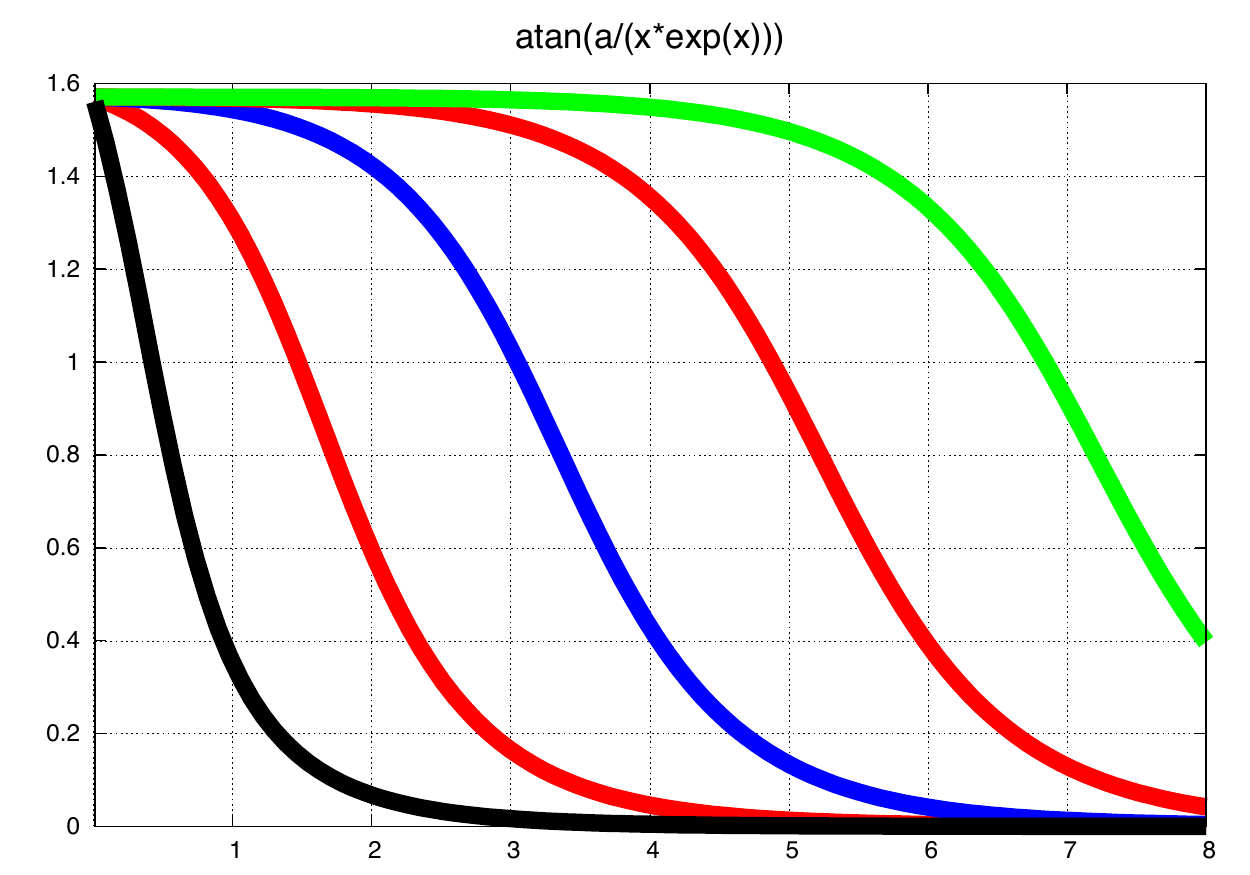}
\end{figure}  \\ 
Function $\arctg\displaystyle\left( \frac a{x\exp(x)} \right)$ has a plateau equal to $\pi/ 2 $ for small values of the parameter $a$.
 Near the value $\pi/ 4$, the function has an inflection point and tends to zero as $x\to\infty$.
Near the point where $x\exp(x) = a$, its argument is equal to one. 

The corresponding solution for $x$ is found from the equation $x = W_0(a)$, where $W_0(a)$ is one of the real branches 
of the Lambert function $W_0(z)$  \cite{Lam, Eul, Dub}.

The behavior of the integrand $\arctg\displaystyle\left( \frac a{x\exp(x)} \right)$ is very close to
behavior of the Heaviside function.
We replace the integrand $\arctg\displaystyle\left( \frac a{x\exp(x)} \right)$ with the Heaviside function,  
which is equal to $\pi/2$ for $x < W_0(a)$, has a jump for $x = W_0(a)$ and vanishes for x greater than $W_0(a)$.
 Thus we get:  
\begin{gather}\label{Havs} 
\int\limits_0^{\infty} \arctg \displaystyle\left(\frac a{x\exp(x)}\right) dx \approx \frac \pi2\int\limits_0^ {W_0(a)} dx = \frac \pi{2} {W_0(a)}
\end{gather}
After differentiation, we get:
\begin{gather}\label{I12}
 I(a) = \frac \pi{2} \frac {W_0(a)}{a(1+W_0(a))} = \frac \pi{2a} \frac {W_0(a)}{1+ W_0(a)}
\end{gather}
Respectively
\begin{gather}\label{Ja}
 J(a) = \frac \pi{2a} \frac {1}{1+W_0(a)}
\end{gather}
The same result can be obtained if we calculate the area of the trapezoid bounded by the abscissa axis, the y-axis, 
the horizontal line  $\pi/2$ and the tangent to the integrand at the inflection point.

In the limit $x \to 0$ expressions (\ref{I12}, \ref{Ja}) coincide with (\ref{i0}, \ref{j0}).  \\
\section{Numerical calculation of integrals.} 
Gaussian quadrature formulas are used to calculate integrals (\ref{sma},\ref{smb}).  
The limits of integration are converted to the interval $[-1, +1] $ using the substitution:  
\begin{gather*} \int\limits_{0}^{+\infty } f(x)\,dx = 
\begin{vmatrix}
x = \displaystyle\frac {1+t}{1-t} \\
dx =  \displaystyle\frac {2 dt}{(1-t)^2} 
\end{vmatrix} =  2 \int\limits_{-1}^{+1} \frac {f\left(\displaystyle\frac {1+t}{1-t}\right)}{(1-t)^2}\,dt = \\
2 \frac {b-a}2 \sum_{i=1}^{n}w_{i}f\left(\frac {1+\xi_{i}}{1-\xi_{i}}\right)\frac 1{(1-\xi_{i})^2} 
\end{gather*} 
that is, the function value must be calculated from the argument $ x = \displaystyle\frac {1+t}{1-t}$,
and the resulting value must be divided by $(1-t)^2 $. The value of $ t$ varies from $-1$ to $+1$.

The nodes and weights of the quadrature formulas are set using the {\bf Gauss\_Legendre()} function (see 
\cite{teuk}[p. 152]), which allows you to install almost any number of integration nodes.
\section{Comparison with numerical calculations }
The  comparison results are presented in three tables: for small values of the parameter $a$, values of the parameter $ 1 \le a \le 10$ and
for large values of $a  > 10$.

In each table the first column contains the value of the parameter $a$, then the numerical value of the integral, then the approximation by the formula
(\ref{Ja}) and the relative error.

The relative error in this table is calculated as the difference between the approximate expression and the numerical value,
referring to the numerical value. That is, the minus sign indicates that the numerical value is greater than the approximating expression.    

\subsection{For small values of the parameter $a$}
For small values of the parameter $a$ ($0 < a < 1.0$), quite acceptable results were obtained:
\begin{table}[ht]
\centering
\begin{tabular}{|c|c|c|c|} \hline
parameter  & Exact    &  Approximate &   Relative   \\
a  & value &    value  & accuracy (\%) \\ \hline
  $1\cdot 10^{-9}$ & 1.5707963066  & 1.5707963252   &  1.18e-06  \\
 $1\cdot 10^{-8}$ & 1.5707961484  & 1.5707963111    & 1.03e-05   \\
 $1\cdot 10^{-7}$   & 1.5707947727 & 1.5707961697    & 8.89e-05   \\
 $1\cdot 10^{-6}$   & 1.5707830885 & 1.5707947560    &   7.42e-04   \\
 $1\cdot 10^{-5}$   & 1.5706869694 & 1.5707806191   &  5.96e-03  \\
 $1\cdot 10^{-4}$  & 1.5699329829  & 1.5706392786    &  4.49e-02 \\
 $5\cdot 10^{-3}$   & 1.5644626729 & 1.5692286650  & 3.04e-01 \\
 $1\cdot 10^{-2}$  & 1.5302197122 & 1.5553956178   & 1.64 \%  \\
 $0.119$  &1.3746503969 & 1.4394118152  &  4.71 \%  \\
 $1.0$  & 1.0031969141 & 1.0023310162  &  -8.63e-02 \% \\ \hline
\end{tabular}
\end{table}  \\  
The relative error tends to zero as $ a \to 0$.  

The largest error value does not exceed five percent (at~$a~=~0.119$). 
\subsection{For the parameter value $ 1 \le a \le 10$}
For the values of parameter $a$ within ($ 1 \le a \le 10$), the results obtained are presented in the table below.  \\
\begin{table}[h]
\centering
\begin{tabular}{|c|c|c|c|} \hline
parameter  & Exact    &  Approximate &   Relative   \\
a  & value &    value  & accuracy (\%) \\ \hline
1.00  & 1.0031969141e+00 & 1.0023310162e+00   & -8.63e-02  \\
2.00  & 8.7427271092e-01  & 8.4788495181e-01    & -3.01e+00  \\
3.00  & 8.0223081578e-01  & 7.6627616508e-01    & -4.48e+00  \\
4.00  & 7.5357260455e-01  & 7.1329545123e-01    & -5.34e+00    \\
5.00  & 7.1747734826e-01  & 6.7511053210e-01    & -5.90e+00   \\
6.00  & 6.8913506420e-01  & 6.4577916569e-01    & -6.29e+00   \\
7.00  & 6.6600973973e-01  & 6.2225892231e-01    & -6.56e+00  \\
8.00  & 6.4661048691e-01  & 6.0280493336e-01    & -6.77e+00 \%     \\
9.00  & 6.2999141037e-01  & 5.8633322110e-01    & -6.92e+00  \%  \\
10.0  & 6.1551743151e-01  & 5.7212904958e-01    & -7.04e+00  \\ \hline
\end{tabular}
\end{table}  
The largest absolute value of the relative error is seven percent.   

All marked results were obtained using the formula (\ref{Ja}).

\subsection{For large values of $ a$}
The table shows the values of the relative error (in percent) for the parameter values
$a$ exceeding $10^{2}$.

\begin{table}[h]
\centering
\begin{tabular}{|c|c|c|c|} \hline
parameter  & Exact    &  Approximate &   Relative   \\
a  & value &    value  & accuracy (\%) \\ \hline
$1\cdot 10^{1}$  & 6.1551743151e-01  & 5.7212904958e-01    & -7.04  \\ 
 $1\cdot 10^{2}$ &  3.83403357e-01  &  3.5816890083e-01  & -6.58 \\
$1\cdot 10^{3}$  &  2.62958980e-01  &  2.5134338355e-01  & -4.41  \\
$1\cdot 10^{4}$  &  1.96425062e-01  &  1.9081944919e-01  & -2.85  \\
$1\cdot 10^{5}$  &  1.55727687e-01  &  1.5273327991e-01  & -1.92    \\
$1\cdot 10^{6}$  &  1.28603919e-01  &  1.2684736369e-01  & -1.36   \\
$1\cdot 10^{7}$  &  1.09333555e-01  &  1.0822372170e-01  & -1.01   \\
$1\cdot 10^{8}$  &  9.49773498e-02  &  9.4234605334e-02  & -7.82e-01    \\
$1\cdot 10^{9}$    &  8.38880237e-02  &  8.3367963716e-02  & -6.19e-01   \\
$1\cdot 10^{10}$  &  7.50753831e-02  &  7.4697789991e-02  & -5.02e-01   \\
$1\cdot 10^{11}$  &  6.79100965e-02  &  6.7627675830e-02  & -4.15e-01  \\ \hline
\end{tabular}
\end{table}  
The relative error in this table is calculated as the difference between the approximate expression and the numeric value 
referring to the numeric value. The minus sign indicates that the numeric value is greater than the approximating expression.
 \begin{figure}[h]
\includegraphics[width=79mm, height=60mm]{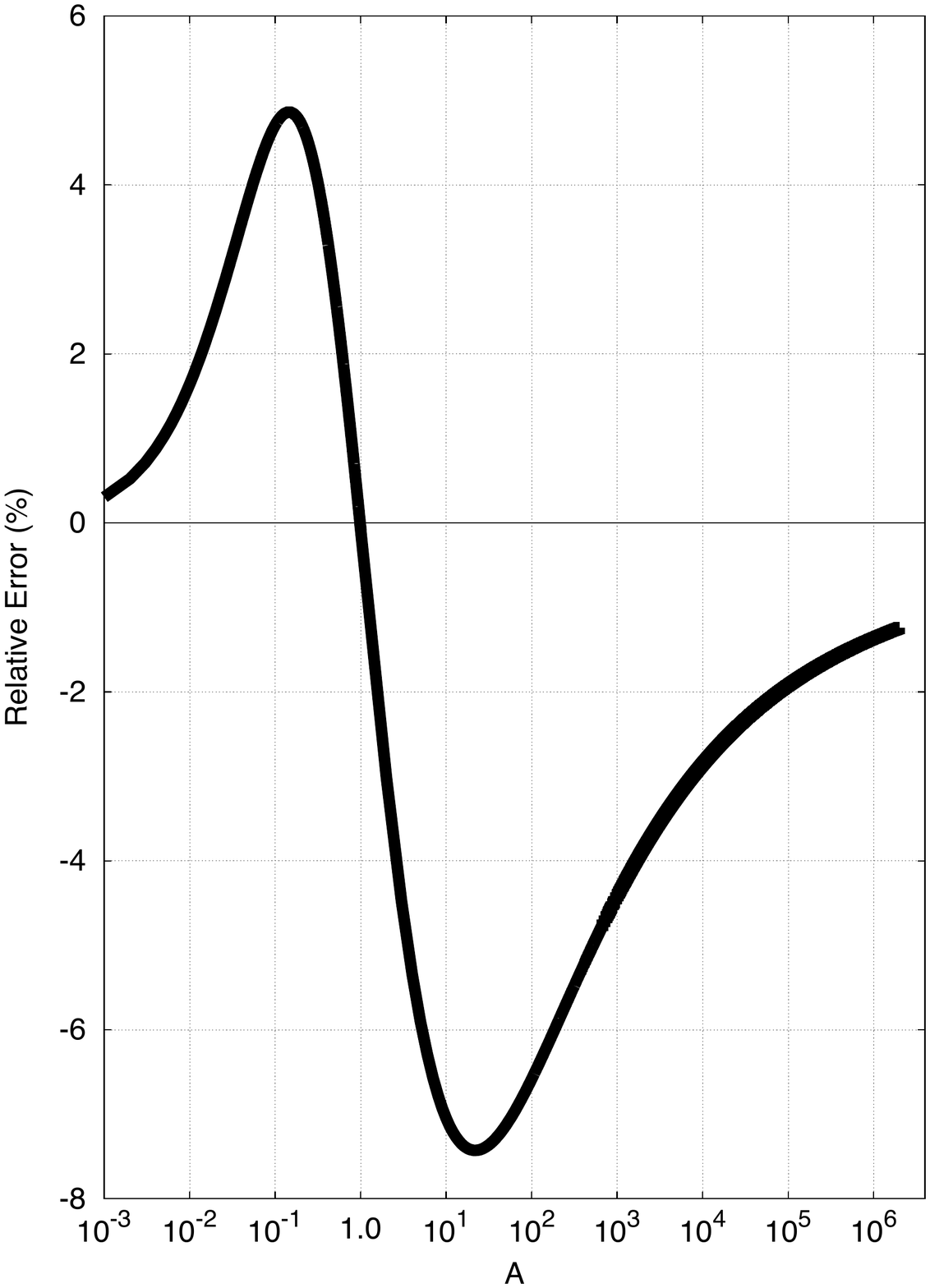}
\end{figure}  \\ 
In the picture below, you can see the relative error behavior on a logarithmic scale.  Near $a = 0.1$, its maximum value is less than $5 \%$.  
  And near $a \approx 22.0$ the maximum error is less than $7\%$ in absolute value.  
  For $a \to 0$ and $a \to \infty$ the relative error tends to zero according to (\ref{Ja}) and (\ref{I12}). 
\section{Program description}
The text of the program consists of the main function and eight functions:  
\begin{itemize}
\item {\bf Gauss\_Legendre} - finds the nodes and weights of the Gauss quadrature formula. In this case, the number of nodes is 24.
\item {\bf gaussGL} - calculates the integral sum over the given nodes and weights.
\item {\bf sect} - divides the integration segment to satisfy the precision of the integral calculation.
\item {\bf topoI} - the function calculates the integrand (\ref{sma}).
\item {\bf topoJ} - the function calculates the integrand (\ref{smb}).
\item {\bf bisection} - finds the value of the Lambert function by the value of the argument $ x = W_0(a)$.
\item {\bf Lambert} - calculates the difference $x\exp(x) - a$. Used by the {\bf bisection} function, which looks for the root of the equation $x\exp(x)=a$.
\item {\bf sign} is a helper function for {\bf bisection}. Specifies the sign of the difference $x\exp(x) - a$.
\end{itemize}
When testing the program, the same trial functions were used as in the previous paper \cite{One}.
\subsection{Main function}
In the text of the main function, all auxiliary print statements are removed. The results of the calculations are printed to the 
LambertTopo.txt file.
\begin{widetext}
\begin{verbatim}
int main()
{  Node = (double*)calloc(NodeNumber, sizeof(double));
   Weight = (double*)calloc(NodeNumber, sizeof(double));
   Gauss_Legendre(-1.0,1.0,Node,Weight,NodeNumber);
   double Xmin = -1.0, rel = 0.0;
   double Xmax = 1.0;
   irecmax = nrec = 60;
   QuadratureRule = gaussGL;
   double int0;
   for(int m = 0; m < nPoints; m++)
    {  for(int k = 0; k < 2; k++)
       {  f1 = fun[k]; 
    Xmin = Left[k];
    Xmax = Right[k];
    estimation = fabs(QuadratureRule(Xmin,Xmax));
    int0 = sect(Xmin,Xmax,estimation); 
    topoIJ[k][m] = A*int0;
    if(m > 0) rel = (topoIJ[k][m] - topoIJ[k][m+1])/M_LN10;
    nofun = unworked = 0;
    irec = 0; 
    double  a = 0.0, c = 25.0;
    double funa, func, root;
    funa = Lambert(a);
    func = Lambert(c);
    root = W0 = bisection(a,c,funa,func,Lambert);
    W[m] = root;
    error[m] = 100*(M_PI_2/(1+W[m]) - topoIJ[1][m])/topoIJ[1][m];
    A += 0.001;
    if(m == 0) fprintf(ass," topoIJ[0]=%8.3lf  topoIJ[1]=%8.3lf  \n",
                topoIJ[0][m],topoIJ[1][m]);
      }
    A = Astart;
    fprintf(ass,"* topoIJ[0]=%8.3lf  topoIJ[1]=%8.3le  \n",
                topoIJ[0][0],topoIJ[1][0]);
    fprintf(ass,"       A            topoI[m]        M_PI_2 - topoI[m] 
                topoJ[m]       M_PI_2/(1+W(a))           W(a)         
                error[m] (%%)     topoI[m]+topoJ[m]  \n");
    for(int m = 0; m < nPoints; m++)
    {fprintf(ass," %8.3le   %18.8le   %18.8le  %20.10le %20.10le 
                %20.10le %20.10le %20.10le\n",A,topoIJ[0][m],
                M_PI_2 - topoIJ[0][m],topoIJ[1][m],M_PI_2/(1+W[m]),
                W[m],error[m],topoIJ[0][m]+topoIJ[1][m]);
        A += 0.001;
    }
    fclose(ass);
    free(Node);
    free(Weight);
    return 1;
}
\end{verbatim} 
\end{widetext}  
During the program execution: we set the values of the nodes and weights of the Gaussian quadrature formulas, 
find the values of the integrals, calculate the value of the Lambert function for a given  $a$, calculate the approximating 
expression and the error value (in percent), and print the results.
\subsection{Recursive adaptive quadrature algorithm}  
The integration algorithm in the program consists of two independent parts: the adaptive function (sect(double, double, 
double)) and the quadrature formula (gaussGL(double, double)). The adaptive function uses a recursive algorithm to 
implement the standard bisection method. This algorithm has been tested many times in previous papers 
\cite{One, BZKINR, BZ, BZh}.

To achieve the required relative accuracy $\varepsilon$ of integration, the integral estimate $\sum_{whole}$  found on the interval 
 [$a_i$,$b_i$] is compared with the sum of $\sum_{left}$ and $\sum_{right}$, which are calculated in the left and right parts of the 
segment [$a_i$,$b_i$]. The comparison rule was chosen in the form:
$$ \sum_{left}+\sum_{right}-\sum_{whole}\leq \varepsilon\cdot \sum_{whole}, $$
where $\sum_{whole}$ denotes the integral sum over the entire integration interval [$a_i$,$b_i$].
The value of $\sum_{whole}$ is accumulated and adjusted at each halving step.

To calculate the integral sums, we use the recursive algorithm for binary division of a segment in half, which has already 
been used in previous works \cite{BZKINR,BZ,BZh}:
\begin{widetext}
\begin{verbatim}
double sect(double x, double X, double int0)
{ if(++irec < irecmax)
{ double sec = (x+X)/2;
    double int1 = QuadratureRule(x,sec);
    double int2 = QuadratureRule(sec,X);
    if((fabs(int1+int2-int0) <= EPS*estimation)) int0 = int1 + int2;
    else int0 = sect(x,sec,int1) + sect(sec,X,int2);
}
else unworked++;
    irec--;
    return int0;
}
\end{verbatim}
\end{widetext}
The choice is due to the  simplicity of this function and good test results on
numerous examples.
\\ 
\subsection{Lambert function evaluation}
To calculate the values of the Lambert function:
$$ x\exp(x) = u , \qquad \Rightarrow \qquad x = W_0(u) $$   
the bisection algorithm, designed as a recursive function, was used \nocite{MCrack}
\begin{widetext}
\begin{verbatim}
double bisection(double& left, double& rigth,  
         double& fun_left, double& fun_rigth, double (*fun)(double&))
{ center = 0.5*(left + rigth);
    if(++irec < irecmax)
    {if(fabs(left - rigth) < 1.0e-14*fabs(center)) return center;
        double fun_center = fun(center);
        if(fabs(fun_left - fun_rigth) < 1.0e-14*fabs(fun_center)) 
                return center;
        if(sign(fun_left) == sign(fun_center)) 
                { left = center ; fun_left = fun_center;  }
        else { rigth = center;  fun_rigth = fun_center;  }
        bisection(left,rigth,fun_left,fun_rigth,fun);
    }
    irec--;
    return center;
}
double Lambert1(double& x)
{     return x*exp(x) - A;   }
\end{verbatim} 
\end{widetext} 

\section{More generali integrals}  
It is easy to see that calculations carried out in Sections \ref{small}  and \ref{approx} can be extended to a 
wider class of integrals.   Consider the following example:
\begin{gather}\label{kain}
K(a) = \frac 1a  \int\limits_0^{\infty}  g(x) \frac  {a}{f^2(x)+a^2}  dx  
 \end{gather} 
 where $f(x)$ is a continuous function that increases monotonically from zero to $\infty$ as the value of 
 $x$ changes within the same limits.  We also assume that function $g(x)$ does not break the convergence 
 of the integral.  Let us introduce the new variable $u=f(x)$ and denote the solution of the equation $f(x)=u$ as $F(u)$, 
 where $F(u)$ is the inverse function of $f(x)$.
 
 After substitution into (\ref{kain}) we get:  
\begin{gather}
K(a) = \frac 1a  \int\limits_0^{\infty}  g(F(u)) \frac  {a}{u^2+a^2}  \frac {du}{f^\prime(F(u))}  
 \end{gather} 
 Let's rewrite the integral as follows:
\begin{gather}
K(a) = \frac \pi{2a}  \int\limits_0^{\infty} \frac {g(F(u))}{f^\prime(F(u))} \frac 2\pi \frac  {a du}{u^2+a^2}     
 \end{gather} 
\subsection{Asymptotic value at $a \to 0$} 
 When $a\to 0$  the following multiplier:   
\begin{gather}
\frac 2\pi \frac {a}{u^2+a^2}  \qquad 
\end{gather} 
 can be considered as delta function normalized from $0$ to $\infty$.  
Which results in:  
\begin{gather}
K(a) = \frac \pi{2a}  \left[ \frac {g(F(u))}{f^\prime(F(u))} \right]_{a=0}     
 \end{gather} 
 In the particular case of $f(x)=x\exp(x)$ and $g(x)=x\exp(x)$ from this expression yields the result obtained in section~\ref{small}.  
\subsection{Approximate expression}  
Under unchanged conditions on the functions $f(x)$ and $g(x)$, consider the same integral:
\begin{gather}\label{cc1} 
K(a) = \int\limits_0^{\infty} \frac  {g(x)}{f^2(x)+a^2}  dx 
\end{gather}
Following the same calculations as in the \ref{approx} section, we integrate (\ref{cc1}) with respect to the parameter {\bf a}:
\begin{gather*}
\int K(a) da = \int \int\limits_0^{\infty} \frac {g(x)}{f(x)} \frac {dx}{1+\left[\displaystyle\frac a{f(x)}\right]^2 } d\left[\frac a{g(x)}\right] = \\
 \int\limits_0^{\infty} \left[\frac {g(x)}{f(x)}\right] \arctg \displaystyle\left(\frac a{f(x)}\right) dx
\end{gather*} 
Since the function $f(x)$ increases monotonically from zero to $\infty$ as the value of $x$ changes within the same limits, the behavior 
of function $\arctg\displaystyle\frac a{f(x)}$ is the same as in the picture in the section \ref{approx}. 
 
Therefore, we can approximately replace $\arctg\displaystyle\frac a{f(x)}$ 
with the Heaviside function. The iflection point satisfies the equation $a = f(x)$.  

This equation can be solved with respect to $x$ as $x = F(a)$, where $F(a)$ is the inverse function to the $f(x)$.  
So we can integrate from 0 to $F(a)$, and the integral reduces to: 
\begin{gather*}
\int K(a) da =  \frac \pi 2\int\limits_0^{F(a)} \left[\frac {g(x)}{f(x)}\right] dx
\end{gather*} 
 where $F(a)$ is the inverse function of $a=f(x)$.  Using the integral differentiation formula with the upper limit 
 depending on the parameter, we find: 
\begin{gather*}
 K(a)  =  \frac \pi 2 \left[\frac {g(x)}{f(x)}\right]_{F(a)}\cdot  \frac {d F(a)}{da}
\end{gather*}  
Let's use the obtained formula to calculate the integral (\ref{Havs}). In this case $F(a) =W_0(a)$ and $g(x) = x\exp(x)$. 
 The derivative of the inverse function is equal to one divided by the derivative of $da/dx$ after $x$ has been 
 replaced by $a$.  Thus:  
\begin{gather*}
 \frac {da}{dx} = (1 + x)\exp(x) \quad \Rightarrow \quad \frac {dx}{da} = \frac 1{(1 + a)\exp(a)}  \\
  K(a)  =   \frac \pi{2a} \frac {W_0(a)}{a (1 + W_0(a))}
 \end{gather*} 
Which is the same as the previously obtained expression (\ref{I12}).  
\section{Conclusion}
The method of finding the approximating expression proposed in the sections \ref{small} and  \ref{approx}   
can be used for not only for (\ref{sma}, \ref{smb})  integrals, but also for many other similar integrals.
For example, all integrals of this type that have a denominator higher than the first order:  
 \begin{gather}\label{sm}
 I(a) = \int\limits_0^{\infty} \frac {x\exp(x)dx}{[(x\exp(x))^2+a^2]^n},    \\  
 J(a) = \int\limits_0^{\infty} \frac {\exp(x) dx}{[(x\exp(x))^2+a^2]^m}   
\end{gather} 
can be represented as derivatives of the appropriate order with respect to the parameter $a$.  

Also, the first term in the denominator does not necessarily have to be $x\exp(x)$. Obviously, this can be any continuous function 
$f(x)$ monotonically increasing from zero to $\infty$ while the value of $x$ changes within the same limits. 
For example, $\sinh(x)$, $x\cosh(x)$ etc.
\section{Acknowledgments}
We expresse our thanks to Dr. V.K.Basenko for stimulating and usefull discussion of the problem.

\end{document}